# *Ex ante* prediction of cascade sizes on networks of agents facing binary outcomes

## Paul Ormerod[1] and Ellie Evans[2]

## September 2010

1. Corresponding author, pormerod@volterra.co.uk;  Volterra Consulting, London, UK; Institute of Hazard, Risk and Resilience, University of Durham, UK;  Extreme Events in Human Society Initiative IIASA Laxenburg, Austria
2. eevans@volterra.co.uk

*Abstract*

*We consider in this paper the potential for* ex ante *prediction of the cascade size in a model of binary choice with externalities (Schelling 1973, Watts 2002). Agents are connected on a network and can be in one of two states of the world, 0 or 1. Initially, all are in state 0 and a small number of seeds are selected at random to switch to state1. A simple threshold rule specifies whether other agents switch subsequently. The cascade size (the percolation) is the proportion of all agents which eventually switches to state 1.*

*We select information on the connectivity of the initial seeds, the connectivity of the agents to which they are connected, the thresholds of these latter agents, and the thresholds of the agents to which these are connected. We obtain results for random, small world and scale –free networks with different network parameters and numbers of initial seeds. The results are robust with respect to these factors.*

*We perform least squares regression of the logit transformation of the cascade size (Hosmer and Lemeshow 1989) on these potential explanatory variables. We find considerable explanatory power for the* ex ante *prediction of cascade sizes. For the random networks, on average 32 per cent of the variance of the cascade sizes is explained, 40 per cent for the small world and 46 per cent for the scale-free.*

*The connectivity variables are hardly ever significant in the regressions, whether relating to the seeds themselves or to the agents connected to the seeds. In contrast, the information on the thresholds of agents contains much more explanatory power. This supports the conjecture of Watts and Dodds (2007.) that large cascades are driven by a small mass of easily influenced agents.*

## 1. Introduction

A wide range of social, economic and popular cultural processes can be characterized as involving 'binary choice with externalities' (Schelling 1973). Agents at any point in time have a choice between two alternatives, and the payoff of an individual is an explicit function of the actions of others. We can usefully apply the same concept to fitness landscapes, where the fitness/survival or otherwise (the binary 'choice') of an agent depends in part upon the fitness of the agent itself, and in part on the state of the world of the agents to which it is connected (for example, Bak and Sneppen 1993 and Solé and Manrubia 1997). A highly relevant and contemporary example is the network which connects financial institutions through their loans to each other as Haldane, an Executive Director at the Bank of England, shows (Haldane 2009). The failure of one bank can lead to a cascade of failures across the system.

Despite enormous resources being devoted to prediction in such situations, such predictions are often woefully inaccurate – market crashes, regime collapses, fads and fashions, and social movements involve significant segments of society but are rarely anticipated. For example, the adoption of innovations (e.g. Arthur 1989, Rogers 2003); diffusion of criminal activity (e.g. Glaeser et. al. 1996) and sociopolitical behaviors (e.g. Hedstrom et al. 2000, Colbaugh and Glass 2009); sales in online markets (e.g. Leskovec et al. 2006); trading in financial markets (e.g. Kirman 1995), and the rise and fall of fads and fashions (e.g., Schelling 1973, Bikhchandani et al. 1998).

Social influence - the impact of the decisions of others on any given agent - decreases the *ex ante* predictability of the ensuing social dynamics. However, Colbaugh et.al. (2010) analyse the data from the artificial cultural market described in Salganik et.al. (2006), and show that these same social forces can *increase* the extent to which the outcome of such a process can be predicted very early in the process.

Watts and Dodds (2007) examine the hypothesis that 'influentials' are important in determining the eventual size of the cascade in such situations, whether it is the choice of a consumer product, adoption of a new technology, forming a particular opinion, or whatever. They conclude that 'large cascades of influence are driven not by influentials, but by a critical mass of easily influenced individuals'.

Here, we consider the extent to which the size of a cascade on a network of agents facing binary choices/outcomes can be predicted purely *ex ante* using a very limited amount of information. We analyse this in the framework of the model developed by Watts (2002), but consider not just random networks, as did the original article, but small world and scale free ones as well. Section 2 outlines the model and section 3 describes the methodology. Section 4 sets out the results.

## 2. The model

We follow the basic approach of Watts (2002). *N* agents are connected on a network. The agents can be in one of two states of the world (0 and 1 for purposes of description), and initially all agents are in state 0. A small number of agents are chosen at random as 'seeds' to switch to state 1.

Each agent is allocated a 'threshold' drawn at random from a uniform distribution on [0, 1]. The agents are therefore heterogeneous. An agent switches from state 0 to state 1 according to the state of the world of the agents to which it is connected. If the proportion of these which are in state 1 exceeds the threshold of the agent, the agent also switches to state 1.

This is a psychologically realistic behavioral rule for exercising choice in many situations. The classic study of Asch (1953), for example, demonstrated the existence of conformity such that the behavior of an agent tended to become more similar to that of the group of which he or she is a member. This could be either because the agent believes the group to have better information than he or she does, or from a desire to conform to group norms.

Yet another potential motivating factor underlying the model is the use of simple copying as a process of social learning. As an efficient strategy of social learning, copying other individuals is often successful (for example, Laland 2004), and is seen right from infancy (Gergely et.al. 2008). The principle is reflected in a recent computer tournament (Rendel et. al. 2010), where entrants submitted strategies specifying how to use both social learning and alternatives such as trial-and-error to acquire adaptive behaviour in a complex multi-player environment. The most successful strategy in the tournament was not (as many expected) a predominance of independent learning supplemented by a relatively minor degree of social learning. In fact, the winning strategy in the tournament relied almost exclusively on copying the successful strategies of others, biased towards copying recent success by 'discounting' older information.

In the context of fitness landscapes, we can usefully think of the threshold as representing the fitness of the agent. State of the world 0, for example, can be thought of as representing existence and state 1 as extinction. Taking the example of a bank, if the proportion of banks to which it is connected which fail exceeds the bank's own fitness, it, too, will fail.

Ormerod and Colbaugh (2006) extend the basic model by allowing the network to evolve in response to agents developing further connections ('alliances') with a view to increasing their fitness. In the basic form described above, each agent carries the same weight in terms of determining the behavior of others. Particularly in networks which influence ideological or religious belief, this may not be a realistic assumption, and Ormerod and Roach (2004, 2008) examine two particular historical episodes where the evidence suggests a weighted scale free network.

However, in this paper we retain the original assumptions of Watts of the network being fixed throughout any individual solution of the model, and each agent carrying the same weight. In addition to the random networks assumed in the original article, however, we also consider small world and scale free networks.

We populate the model with 1000 agents and connect the agents on a network. We obtain 1000 separate solutions for each parameterization of the network, with each solution having its own unique network constructed from the relevant parameters. A small number of agents are drawn at random as seeds to be the initial converts to state 1. The model evolves in a series of steps, in each of which all agents in state 0 decide whether or not to convert to state 1. The solution halts when no single agent switches in the step.

### 3. Methodology

#### 3.1 Choice of potential explanatory factors

We denote by $\pi(i)$ the size of the percolation in the *i'th* solution of the model, and $\pi$ as the vector of the 1,000 solutions. The percolation is the proportion of the total number of agents which switch to state 1.

We considered a number of variables in order to investigate the extent to which the size of the percolation could be explained by a limited amount of information available *ex ante* before a run of the model is carried out. In terms of the seeded agents, we calculated both the average and the maximum degree of the seeds. We expect a positive correlation between each of these and the size of the percolation.

We examined the characteristics of the agents connected to the seeds and extracted the minimum, mean and maximum threshold of these agents. The higher are the minimum and mean thresholds of these agents, the lower we expect the percolation to be, whilst we expect a positive correlation between the value of the maximum threshold and the percolation.

We also created a variable we designate as 'below'. This is the number of agents connected to a seed which have a threshold below some critical value. In other words, they are agents which have a high propensity to switch from state 0 to state 1.

For the random networks, this critical value has a natural definition. Denoting by $p$ the probability of connecting any pair of agents, on average each agent will have $v = pN$ connections, where $N$ is the total number of agents. So on average an agent which is connected to a seed and which has a threshold less than $v$ will be certain to convert. The small world networks also have a rather natural definition of the critical value, and we use $1/n$ where $n$ is the number of immediate neighbours of each agent. We ignore any additional connections formed by re-wiring in this definition. The definition of the critical value is less obvious in the case of scale-free networks, given that the typical (average) number of connections of an agent has much less meaning in this kind of network, and indeed the population mean may not even exist. We experimented with a range of values between 0.1 and 0.25 with which to define those agents designated as being 'below' a critical value of their thresholds, and the results were robust with respect to reasonable variations in this range.

To repeat, the variable 'below' is the number of agents connected to seeds which have low thresholds and are therefore particularly likely to switch states of the world. We expect that the higher this is, the higher will be the percolation.

We also examined after the first step of the solution of the model, those agents which actually switch in response to the seeds. We extract information on the minimum, mean and maximum thresholds of agents connected to the agents which switch in the first step of the solution. Again, we expect a negative correlation between the minimum and mean values and the size of the percolation, and a positive one with the maximum value.

So we have 12 candidate variables with which to try to account for the size of the percolation in any given solution of the model. Some relate to the connectivity of the seeds, some to the propensity to switch of those agents connected to the seeds, and some to the propensity to switch of those agents connected to agents which switch in the first step of the solution.

We also carried out the analysis omitting the information on the three variables which relate to the thresholds of agents connected to the ones which switch in the first step of the solution. In a real world situation, the difficulty of obtaining accurate information about the agents obviously increases the further removed from the initial disturbance (the seeds switching to state 1) any given agent is. So we also analyse the predictability of percolations without these 'second-phase' agents.

*3.2    Statistical procedure*

Initially, we carried out univariate screening of the impact of each potential explanatory variable (Hosmer and Lemeshow 1989). We tested the null hypothesis that the distribution of $\pi$ for the solutions in which the level of the explanatory variable is less than its 1st quartile value is the same as the distribution of $\pi$ for the solutions in which the level of the explanatory variable is greater than its 3rd quartile value.

Given that the data on both $\pi$ and most of the potential explanatory variables is distinctly non-Gaussian, we use the Anderson-Darling test (Anderson and Darling 1952), which is known to have greater power in such circumstances than the more widely used Kolmogorov-Smirnov.  On standard statistical criteria, the null hypothesis is deemed to be rejected at p-values < 0.05, but we apply the more relaxed criterion of a p-value of 0.10 for rejection.  Heuristically, the higher the p-value criterion at which the null hypothesis is deemed to be rejected, the more similar the two distributions are.  We choose a p-value of 0.10 because do not want to leave out variables in the subsequent analysis which might be important (Mickey and Greenland 1989).

We retained as candidate variables for explanation those in which the null hypothesis is rejected at a p-value of < 0.10.  For p-values > 0.10, the distribution of $\pi$ when the potential explanatory variable takes values less than its 1st quartile can be regarded as the same as when it takes values greater than its 3rd quartile, suggesting the variable in question does not have an effect on percolation.

The variable to be explained, $\pi$, is bounded in [0,1], and so we formed the logit transformation (Hosmer and Lemeshow op.cit.)  $\log(\pi/1-\pi)$ to use as the dependent variable in multi-variate regression.

We performed least-squares regression of this logit transformation on all the potential explanatory variables identified by the screening process above.  We then eliminated one-by-one the variable with the highest p-value (provided the p-value > 0.05) in the test of the null hypothesis that the value of its coefficient in the regression is zero (the lowest value on the standard t-test).  We used as our preferred equation the one in which for no single coefficient is the null hypothesis rejected at a p-value > 0.05.

The next step was to bootstrap the preferred equation 1,000 times, to check the assumption of normality in the standard errors on the individual coefficients (Davison and Hinckley 1997).  In each case, the bootstrap procedure indicated that the least-squares results were acceptable i.e. little or no bias was found in the estimated coefficients and their associated standard errors.

Finally, we carried out screening for potential non-linearities in the impact of each of the explanatory factors using the local linear regression (Cleveland and Devlin 1988) procedure in the statistical package S-Plus.  In general, the assumption of linearity was found to be reasonable, and there were no marked non-linearities.

4. **Results**

We investigated a reasonably wide range of parameters of the three types of networks - random, small world and scale free – and used different numbers of initial seeds.  In terms of seeds, we examined solutions with 5, 10 and 20 seeds out of the total of 1000 agents.  For the random networks, we examined values of *p* between 0.005 and 0.04, and for the small world the number of immediate neighbours of each agent varied between 5 and 20, with the probability of an additional re-wired connection being between 0.01 and 0.11.  For the scale-free, the networks were formed by by varying the initial number of connected vertices between 2 and 8, and the expected number of undirected links each period between 1 and 5.

The results were robust with respect to different parameter values for the networks in the sense that the potential variables fall into three classes across the range of networks and number of initial seeds examined: those which are either always or almost always significant; those which are significant intermittently; those which are never or almost never significant.

We present illustrative results of the regressions below in Table 1 for the three kinds of network. Several 'templates' of the distribution of $\pi$, the size of the percolation, are represented in each. In some, there is a low median value of the percolation and a low maximum, a low median and a 'medium sized' maximum, and in others a low median and a high maximum, corresponding to a percolation of close to global size across the population.  In others, higher values of the median are associated with 'medium' and high values of the maximum.   The distribution of $\pi$ in every single parameterisation of the network is non-Gaussian and heavily right-skewed.

Table 1 shows the results, which are entirely typical, for 12 parameterisations of the random network, 12 of the small world, and 18 of the scale free

**Table 1      Number of times the variable has a coefficient statistically significant from zero at p-value < 0.05 in the preferred multiple regression**

**Network type**

| Variable | Random | Small world | Scale free |
|---|---|---|---|
| Min threshold of agents connected to seeds | 2 | 0 | 13 |
| Mean            " | 9 | 12 | 8 |
| Max             " | 0 | 0 | 2 |
| Number of agents connected to seeds with threshold below 'critical value' | 12 | 12 | 18 |
| Mean degree of seeds | 5 | 0 | 7 |
| Max degree of seeds | 0 | 0 | 0 |
| Min degree of agents connected to seeds | 0 | 0 | 0 |
| Mean          " | 0 | 0 | 0 |
| Max           " | 0 | 0 | 0 |
| Min threshold of agents connected to the agents connected to the seeds | 11 | 7 | 18 |
| Mean          " | 11 | 12 | 18 |
| Max           " | 6 | 8 | 18 |

**Note:** *These are the results for 12 random networks, 12 small world, 18 scale free; in each case the dependent variable is the proportion of agents which eventually switch to state 1 in each of 1,000 separate solutions of the model*

The connectivity variables were hardly ever significant in the regressions, whether relating to the seeds themselves or to the agents connected to the seeds. So even in the scale free regressions, for example, choosing as a seed one of the highly connected hubs of the network does not have a significant influence on the eventual size of the percolation.

In contrast, the information on the thresholds of agents contained much more explanatory power. In particular, the variable measuring the number of agents connected to seeds with a threshold below some (low) 'critical value' is significantly different from zero in every single regression. This supports the finding of Watts and Dodds (op.cit.) that large cascades are driven by a small mass of easily influenced agents.

The mean threshold of agents connected to seeds is significant in some three-quarters of the regressions, even though the range of this variable can be small especially when 20 seeds are used in the model solutions.

The thresholds of agents connected to agents connected to seeds – one step removed as it were from the initial starting point – are also for the most part significant. It would in principle be possible to extend this even further and look at, for example, the thresholds of agents connected to agents who themselves are connected to seeds, but clearly in any practical context acquiring reasonably accurate information on factors such as this would become more difficult the further removed from the initial conditions that we look.

Table 2 presents standard information on the degree of explanatory power of the regressions reported in Table 1. We do this both for the entire set of variables, and for the set which excludes the three 'once removed' variables.

**Table 2**      **Explanatory power of the regressions in Table 1: $R^2$ statistics of the regressions**

|  | full set of variables | | | excluding the 3 'once removed' variables | | |
|---|---|---|---|---|---|---|
|  | average | min | max | average | min | max |
| **Random** | 0.32 | 0.23 | 0.42 | 0.22 | 0.13 | 0.34 |
| **Small world** | 0.40 | 0.30 | 0.52 | 0.33 | 0.21 | 0.43 |
| **Scale free** | 0.46 | 0.39 | 0.59 | 0.35 | 0.28 | 0.42 |

Even with a very small amount of information, a reasonably powerful explanation of the eventual size of the percolation can be given, especially for the two more structured networks compared to the purely random one. With the scale free networks, for example, on average 46 per cent of the total variance across the 1000 percolations generated by each parameterization of the network is explained, and in one particular case as much as 59 per cent.

In general, the required information relates to the thresholds of agents either connected directly to seeds, or to agents connected to those connected to seeds. In practice, of course, such information might not be completely straightforward to obtain. But given the explanatory power of the regressions, there is a good incentive to acquire it, both for example in a marketing context when launching a new brand and in the context of financial regulators assessing the potential for a cascade of failures across institutions. It is not so much the vulnerability of any given institution to failure, for example, rather it is the vulnerability of the financial institutions with which a failing institution is most directly connected.